\documentstyle{aipproc}
\newcommand{\hepex}[1]{(hep-ex/#1)}
\newcommand{\hepph}[1]{(hep-ph/#1)}
\def\prl#1#2#3{\frenchspacing{\it Phys. Rev. Lett. }{\bf #1}, #2 (19#3)}
\def\pr#1#2#3{\frenchspacing{\it Phys. Rev. D}{\bf #1}, #2 (19#3)}
\def\pl#1#2#3{\frenchspacing{\it Phys. Lett. }{\bf #1}, #2 (19#3)}
\def\np#1#2#3{\frenchspacing{\it Nucl. Phys. }{\bf #1}, #2 (19#3)}
\def\rmp#1#2#3{\frenchspacing{\it Rev. Mod. Phys. }{\bf #1}, #2 (19#3)}
\def\epj#1#2#3{\frenchspacing{\it Euro. Phys. J. }{\bf C#1}, #2 (19#3)}

\def\ib#1#2#3{{\it ibid. }{\bf #1}, #2 (19#3)}
\def\yadfiz#1#2#3#4#5#6{{\it Yad. Fiz. }{\bf #1}, #2 (19#3) [English translation: 
         {\it Sov. J. Nucl. Phys. }{\bf #4}, #5 (19#6)]}
\def\nat#1#2#3{{\it Nature (London) }{\bf #1}, #2 (19#3)}

\newcommand{\Imag}[1]{\mathop{\rm Im}(#1)}
\newcommand{\Real}[1]{\mathop{\rm Re}(#1)}

\def\ket#1{| #1\rangle}

\newcommand{\etal}{{\em et al.}}
\newcommand{\ie}{{\em i.e.}}

\newcommand{\gevcc}{\hbox{ GeV}\!/\!c^2}

\newcommand{\gevc}{\hbox{ GeV}\!/\!c}

\newcommand{\mev}{\hbox{ MeV}}

\newcommand{\mevcc}{\hbox{ MeV}\!/\!c^2}

\newcommand{\kev}{\hbox{ keV}}
\newcommand{\km}{\hbox{ km}}
\newcommand{\fermi}{\hbox{ fm}}

\newcommand{\mb}{\hbox{ mb}}

\newcommand{\fb}{\hbox{ fb}}
\newcommand{\ps}{\hbox{ ps}}

\def\ltap{\mathop{\raisebox{-.4ex}{\rlap{$\sim$}} 
\raisebox{.4ex}{$<$}}}
\def\gtap{\mathop{\raisebox{-.4ex}{\rlap{$\sim$}} 
\raisebox{.4ex}{$>$}}}
\def\eqn#1{(\ref{#1})}
\newcommand{\hq}{\textit{HQ98}}
\newcommand{\cfrac}[2]{\textstyle \frac{#1}{#2}}
\def\bentarrow{\:\raisebox{1.1ex}{\rlap{$\vert$}}\!\rightarrow}

\def\dk#1#2#3{
	\begin{displaymath}
	\begin{array}{r c l}
	#1 & \rightarrow & #2 \\
	 & & \bentarrow #3
	\end{array}
	\end{displaymath}
		}

\def\dkp#1#2#3#4{
	\begin{displaymath}
	\begin{array}{r c l}
	#1 & \rightarrow & #2#3 \\
	 & & \phantom{\; #2}\bentarrow #4
	\end{array}
	\end{displaymath}
		}

\begin{document}
\begin{flushright}
\rule{0pt}{36pt} FERMILAB--CONF--98/390--T 
\vspace{-30pt}
\end{flushright}
\title{Perspectives on Heavy Quark 98}

\author{Chris Quigg}
\address{Fermi National Accelerator Laboratory\thanks{Fermilab is 
operated by Universities Research Association Inc.\ under Contract 
No.\ DE-AC02-76CH03000 with the United States Department of Energy.}\\
P.O. Box 500, Batavia, Illinois 60510 USA \\ E-mail: \textsf{quigg@fnal.gov}}

\lefthead{Perspectives on Heavy Quark 98 \phantom{FERMILAB--CONF--98/390--T}}
\righthead{\phantom{Perspectives on Heavy Quark 98} FERMILAB--CONF--98/390--T}
\maketitle

\begin{abstract}
I summarize and comment upon some highlights of \hq, the Workshop on 
Heavy Quarks (strange, charm, and beauty) at Fixed Target.
\end{abstract}

\section*{Historical Prologue}
Half a century has passed since George Rochester and Clifford Butler 
announced their discovery of ``vee particles,'' penetrating products 
of cosmic-ray showers that proved to be $K$ mesons, the first strange 
particles \cite{randb}.  Through the years, it is striking how 
thoroughly the study of heavy flavors has defined our progress toward 
an elegant and comprehensive picture of the fundamental constituents 
and their interactions.  Understanding heavy flavors 
has been essential to understanding the ordinary stuff of everyday 
matter.

To kick off \hq, we had the pleasure of hearing reminiscences from 
Lincoln Wolfenstein \cite{wolf}, Jon Rosner \cite{jlr}, and Tony 
Sanda \cite{tony} on the beginnings of our understanding of 
strangeness, charm, and beauty.  They offered interesting lessons in 
where we have been, and where we hope to go.

Not all history lies so far in the past.  \hq\ weekend witnessed an 
important event for the future of Fermilab and of heavy-quark 
physics.  At 17:21 on Saturday, October 10, circulating beam was 
established for the first time in the Main Injector.  This new 
element in Fermilab's cascade of accelerators, a proton synchrotron 
precisely $\pi\km$ around, will play a double role as injector to the 
Tevatron and as the high-intensity source for a new 120-GeV 
fixed-target program at Fermilab.  On behalf of the participants in 
\hq, it is my pleasure to salute our colleagues for this fine 
achievement, and to wish them continued success in commissioning our 
newest accelerator.

\section*{\boldmath{$B\hbox{ -- }\bar{B}$} mixing and related topics}

\subsection*{Mixing Phenomenology and Experiment}

I gave a similar conference summary at \textit{Heavy Flavors'87} at 
Stanford \cite{stanford87}, so it was natural to look through my transparencies from 
that meeting while preparing this talk.  Although I agree with the many 
speakers here at \hq\ who have said that we are just at the beginning 
of the study of this or the serious study of that, what struck me 
most was the very dramatic progress we have made in nearly every aspect of 
heavy-quark physics.

To take a prominent example, in 1987 we were just digesting the first 
evidence for particle-antiparticle mixing in the neutral $B$ mesons.  
For some time, there had been provocative indications from the UA1 
experiment in the form of an excess of same-sign dimuon events over 
what can be accounted for in the absence of mixing \cite{ua1mix}.  
Because of theoretical prejudice that $B_s-\bar{B}_s$ mixing might be 
large, and because of published upper bounds on the rate of 
$B_d-\bar{B}_d$ mixing, this result was taken as the scent of 
$B_s-\bar{B}_s$ mixing.  Since this interpretation relied on 
simulations, and UA1 had not reconstructed any $B$ mesons, the case 
for mixing was not proved.

The needed proof was supplied by the \textsc{argus} Collaboration working
at the DORIS storage ring \cite{argus}. The demonstration that $B_d-\bar{B}_d$
mixing takes place came in the form of a single (nearly) reconstructed
$B^0B^0$ event produced in the chain
\dkp{e^+e^-\to\Upsilon(4S)}{B^0}{\bar{B}^0}{B^0\; .}
The two neutral $B^0$s, which must be nonstrange because the $B_s$
cannot be pair-produced at the $\Upsilon(4S)$, were identified in
the decay chains
\begin{equation}
\begin{array}{r c l}
	B^0 & \to & D^{*-}\mu^+\nu \\
	& & \bentarrow \pi^-\bar{D}^0 \\
	& & \phantom{\bentarrow \pi^-}\bentarrow K^+\pi^-
\end{array}
\end{equation}
and
\begin{equation}
\begin{array}{r c l}
	B^0 & \to & D^{*-}\mu^+\nu \\
	& & \bentarrow \pi^0\bar{D}^- \\
	& & \phantom{\bentarrow}\:\raisebox{1.3ex}{\rlap{$\vert$}}\raisebox{-0.5ex}{$\vert$}
	\phantom{\pi^0}\!\bentarrow K^+\pi^-\pi^-\; , \\
	& & \phantom{\bentarrow}\bentarrow \gamma\gamma
\end{array}
\end{equation}
both fully reconstructed, except for the undetected neutrinos. Inspired
by this event, the \textsc{argus} experimenters carried out two statistical
analyses using dilepton events or incompletely reconstructed 
$B^0\rightarrow D^*\ell\nu$ events to determine the degree of $B_d-\bar{B}_d$
mixing. They estimated $x_{d} \equiv \Delta M_B/\Gamma_B\simeq 0.7$ 
(where $\Gamma_{B}$ is the average lifetime of the heavy and light 
$B^{0}$ states, to be compared with $\Delta M_K/\Gamma_K\simeq 0.5$.

At \hq, we have seen two lovely examples of \textit{time-dependent} 
$B^{0}\hbox{ -- }\bar{B}^{0}$ oscillations in the talks by Kevin 
Pitts \cite{pitts}, representing CDF, and Achille Stocchi 
\cite{stocchi}, representing the four LEP collaborations.  These 
plots, which are based on thousands of clean events, will take their 
place in textbooks alongside the classic plots of the time evolution 
of $K^{0}\hbox{ -- }\bar{K}^{0}$ oscillations.  They represent 
phenomenal progress since the discovery of $B^{0}\hbox{ -- }\bar{B}^{0}$ 
mixing in 1987.  And this is not all.  Through the combined efforts of 
\textsc{aleph}, \textsc{delphi}, \textsc{opal}, and L3 at LEP, CDF at 
the Tevatron, and SLD at the Stanford Linear Collider, we now can 
quote a very precise world average for the mass difference \cite{pitts}
\begin{equation}
	\Delta M_{B} = 0.475 \pm 0.010_{\mathrm{stat}} \pm 
	0.014_{\mathrm{sys}}\ps^{-1}.
	\label{eq:delmd}
\end{equation}
so that $x_{d} \approx 0.74$.

This much is solid achievement, but a great deal more is in the works.  
We expect the frequency of $B_{s}\hbox{ -- }\bar{B}_{s}$ oscillations 
to be much more rapid than that of $B^{0}\hbox{ -- }\bar{B}^{0}$.  The 
LEP experiments plus CDF and SLD can now set a lower bound of $\Delta 
M_{B_{s}} > 12.4\ps^{-1}$ \cite{bosco}, which implies $x_{s} \equiv 
\Delta M_{B_{s}}/\Gamma_{B_{s}} \gtap 18.5$.  An observation of 
$B_{s}\hbox{ -- }\bar{B}_{s}$ mixing would fix the ratio of the 
quark-mixing matrix elements $V_{td}$ and $V_{ts}$.  While the LEP 
experiments continue to press their analyses, the greatest reach in 
the near future will come from CDF and D\O, using the $2\fb^{-1}$ of data 
each experiment will accumulate in Run II of the Tevatron Collider.  
For the baseline detector, CDF anticipates a reach in the range 
$x_{s} \approx 30\hbox{ - }40$; additional upgrades could extend the 
reach to $x_{s} \approx 55\hbox{ - }65$ \cite{pitts}.

In the standard electroweak theory, the dominant contributions to $B-\bar{B}$
mixing come from box diagrams involving loops of $W$ bosons and quarks,
most importantly top quarks. These lead to expressions for the mixing
parameters,
\begin{equation}
	x_{d} = \frac{\Delta M_{B}}{\Gamma_{B}} \propto f_{B_d}^2|V_{td}^{*}V_{tb}|^2
B_{B_d} \tau_{B_d} m_t^2 
\end{equation}
and
\begin{equation}
	x_{s} = \frac{\Delta M_{B_{s}}}{\Gamma_{B_{s}}} \propto 
	f_{B_s}^2|V_{ts}^{*}V_{tb}|^2
B_{B_s} \tau_{B_s} m_t^2 \; ,
\end{equation}
that contain many parameters.  I think the worst moment in my career 
as a summary speaker came during that talk at \textit{Heavy Flavors - 
'87.}  When I flashed these formulas on the screen, I suddenly became 
aware that I could pronounce the name of every parameter, but I didn't 
know the value of a single one!  Our ignorance in 1987 ranged from 
the uncertain relationship between quark matrix elements
and hadronic matrix elements subsumed in the infamous $B$ parameters 
to the mass of the top quark and the $B$-meson lifetimes.

I'm very happy that a decade of progress means I do not have to 
relive that unsettling moment.  Harry Cheung's review of charm and 
beauty lifetimes \cite{harryc} presents us with wonderfully precise 
values for $\tau_{B_{d}} = (1.556 \pm 0.027)\ps$ and $\tau_{B_{s}} = 
(1.489 \pm 0.058)\ps$.  The top mass, which was entirely unknown in 
1987, is now known to very impressive precision.  My informal 
average of the latest results from CDF \cite{cdftop} and D\O\ 
\cite{d0top} yields $m_{t} = (173.8 \pm 4.8)\gevcc$.  Andreas Kronfeld 
reported on the development of lattice QCD calculations of the 
pseudoscalar decay constants and related parameters \cite{andreas}.  
The study of heavy ($b$ and $c$) quarks on the lattice, which 
coincidentally began in 1987, is now a mature subject.  I think it is 
fair to say that almost definitive calculations of $f_{B}$ and 
$f_{B_{s}}$ (as well as $f_{D}$ and $f_{D_{s}}$) are in hand, and 
that convergence on the $B_{B}$ parameters is on the horizon.  It would be 
incautious of me to record ``best values,'' but I think it is useful 
to quote \textit{representative values} drawn from Kronfeld's 
compilation: $f_{B} \approx 165 \pm 15\mev$, $f_{B_{s}} \approx 188 
\pm 15\mev$, $f_{D} \approx 195 \pm 15\mev$, and $f_{D_{s}} \approx 
220 \pm 15\mev$, not including the estimated effects of quark loops, 
which are thought to increase the values about 10\%.  Our best 
experimental test of these calculations comes from the purely leptonic 
decays $D_{s} \rightarrow \mu\nu\hbox{ and }\tau\nu$, which yield a 
world average value, $f_{D_{s}} = 254 \pm 31\mev$ \cite{stocchi}.  
This is encouragingly close to the calculated value adjusted for quark 
loops.  The ``bag factors'' $B_{B}$ are not in such settled condition; 
the scatter among calculated values still exceeds the uncertainties 
attributed to the calculations.  More work is needed, but I do hope 
that convergence on reliable values is near.  I am encouraged to note 
that the suitably defined bag factor for the neutral kaon system does 
seem to have converged to a value near 0.62 (see Ref.\ \cite{andreas} 
for the precise definition), with little sensitivity to the omission 
of light-quark loops.

The remaining quantities, the quark-mixing matrix elements, are known 
to reasonable precision if we assume the three-generation picture and 
invoke unitarity of the CKM matrix to fix their values.  (The Particle 
Data Group advises $|V_{td}| = 0.004\hbox{ to }0.013$, $|V_{ts}| = 
0.035\hbox{ to }0.042$, and $|V_{tb}| = 0.9991\hbox{ to }0.9994$ 
\cite{pdg}.)  But if we demand a measurement, it is from the study of 
$B\hbox{ -- }\bar{B}$ oscillations that we get our best information 
about $|V_{td}|$ and $|V_{ts}|$.  Single-top production in $\bar{p}p$ 
collisions will give us our first measurement of $|V_{tb}|$.

\subsection*{{\boldmath$CP$} Violation in the {\boldmath$B$} System}
One of our very important near-term goals is the observation and 
detailed study of $CP$ violation in $B$ meson decays.  If the observed 
$CP$ violation in the neutral kaon system indeed arises from the phase 
in the Cabibbo--Kobayashi--Maskawa quark mixing matrix, then the 
manifestations of $CP$ violation in $B$ decays will be rich and 
informative.  Distinguishing $B^{0}$ from $\bar{B}^{0}$ by means of a 
same-side--tagging technique, the CDF Collaboration has performed a 
first measurement of the asymmetry
\begin{equation}
	{\mathcal A} = \frac{N(\bar{B}^{0} \rightarrow \psi K_{S})
	- N(B^{0} \rightarrow \psi K_{S})}
	{N(\bar{B}^{0} \rightarrow \psi K_{S})
	+ N(B^{0} \rightarrow \psi K_{S})},
	\label{eq:CPasym}
\end{equation}
using 200 events in which both muons from the decay $\psi \rightarrow 
\mu^{+}\mu^{-}$ are reconstructed in the silicon vertex detector.  In 
the standard model, this time-varying $CP$-violating asymmetry is given by
\begin{equation}
	{\mathcal A}(t) = \sin(\Delta M_{B}t) \cdot \sin{2\beta},
	\label{eq:SMCPasym}
\end{equation}
where $\beta$ is the angle in the complex plane between $V_{td}$ and 
$V^{*}_{cb}$.  The CDF measurement \cite{pitts,cdfcp},
\begin{equation}
	\sin{2\beta} = 1.8 \pm 1.1_{\mathrm{stat}} \pm 0.3_{\mathrm{sys}},
	\label{eq:CDFasym}
\end{equation}
is statistics limited.  The dominant contribution to the systematic 
error comes from the uncertainty in the dilution factor.  While it is 
more in the nature of a dress rehearsal than an informative 
measurement, this CDF exercise is an important step on the path toward 
the discovery of $CP$ violation in the $B$ system \cite{opalcp}.

The search for $CP$-violating effects in the $B$ system will soon 
begin in earnest.  The fixed-target experiment \textsc{hera--b} has 
begun to take data, as we heard from B.  Schwingenheuer \cite{herab}.  
Commissioning is well under way at the SLAC $B$ factory, and the 
\textsc{BaBar} experiment described by G.  Bonneaud \cite{babar} is 
approaching completion, with first data expected in 1999.  Also in 
1999, we expect the first run of \textsc{belle} at the KEK $B$ factory 
\cite{belle}.  The following year, the upgraded CDF and D\O\ detectors 
will profit from the greatly enhanced luminosity that the Main 
Injector will bring to the Tevatron Collider \cite{pitts}.  Farther in 
the future are \textsc{lhc}$b$ \cite{lhcb} and the BTeV proposal at 
Fermilab \cite{btev}.  A general survey of the promise of forthcoming 
experiments was presented at \hq\ by Marina Artuso \cite{artuso}.

While the new experiments prepare themselves for serious data-taking, 
the highly successful Cornell Electron Storage Ring, which we may 
regard as a stationary center-of-momentum $B$ factory, continues to 
produce a rich harvest of physics results in the upgraded \textsc{cleo} 
detector.  Among the \textsc{cleo} results presented at \hq, those 
reported by Peter Gaidarev on rare decays \cite{pgcleo} are of special 
interest to the search for $CP$ violation.  The \textsc{cleo} 
measurements of the branching fractions for $B^{0} \rightarrow K\pi, 
\pi\pi, \hbox{ and }KK$ will inform the search for, and interpretation 
of, a $CP$ asymmetry in the decays $(B^{0},\bar{B}^{0})\rightarrow 
\pi^{+}\pi^{-}$.  The current values are $B(B^{0} \rightarrow 
K^{\pm}\pi^{\mp}) = (1.4 \pm 0.3_{\mathrm{stat}} \pm 
0.2_{\mathrm{sys}}) \times 10^{-5}$ and $B(B^{0} \rightarrow 
\pi^{+}\pi^{-}) < 0.8 \times 10^{-5}$ at 90\% CL.  The small 
$\pi^{+}\pi^{-}$ branching fraction complicates the program to extract 
the angle $\alpha$ between $V^{*}_{ub}$ and $V_{td}$.

Before leaving the \textsc{cleo} data on rare decays, let us note that 
the new measurement of the inclusive $b \rightarrow s\gamma$ branching 
fraction,
\begin{equation}
	B(b \rightarrow s\gamma) = (3.15 \pm 0.35_{\mathrm{stat}} \pm 
	0.32_{\mathrm{sys}} \pm 0.26_{\mathrm{model}}) \times 10^{-4},
	\label{eq:bsg}
\end{equation}
is in good agreement with standard-model expectations, and limits the 
phase space of models for new physics.  

\subsection*{Baryogenesis}
Together with the existence of fundamental processes that violate 
baryon number and a departure from thermal equilibrium during the 
epoch in which baryon-number--violating processes were important, 
microscopic $CP$ violation is a necessary condition for generatng a 
nonvanishing baryon number from an initially symmetrical universe.  As 
Peter Arnold reviewed in his talk at \hq\ \cite{pba}, the $CP$ 
violation we attribute to a phase in the quark mixing matrix does not 
appear capable of generating a baryon-to-photon ratio nearly as large 
as the value
\begin{equation}
	n_{B}/n_{\gamma} \approx 10^{-9 \pm 1}
	\label{eq:barphot}
\end{equation}
inferred from astronomical observations.  Nevertheless, we hope that 
lessons learned from the study of $CP$-violating phenomena in the 
domain of heavy quarks will inform our eventual understanding of the 
baryon number of the universe.  One of the most active areas of recent 
theoretical work has been to elaborate the possibility that 
baryogenesis occurs on the scale of electroweak symmetry breaking.  In 
this scenario, it is not possible to generate a large enough 
baryon-to-photon ratio in the minimal electroweak theory with a single 
Higgs doublet.  The feat can be accomplished in a supersymmetric 
theory, but only if the lightest Higgs boson is very light, $M_{h} 
\ltap 105\gevcc$, and the stop squark weighs no more than the top quark 
\cite{marcela}.  Under these conditions, both $h$ and $\tilde{t}$ 
should be accessible soon at LEP200 and the Tevatron Collider, and we 
can expect departures from the standard-model $CP$ phenomenology in 
the $B$ mesons.

\section*{Heavy-Flavor Spectroscopy}
\subsection*{Excited Mesons}
For many years, the principal focus of charm and beauty physics has 
been on the weak decays of states stable against strong or 
electromagnetic decay.  From these, we have learned important lessons 
about the structure of the weak charged current and about the 
interplay between the strong and weak interactions.  Over the past 
five years or so, the study of excited states---especially excited 
meson resonances---has taken on a new interest, as high-statistics 
experiments with excellent mass resolution have attained maturity.  
We now know a good deal about the meson states $c\bar{q}$, 
$c\bar{s}$, $b\bar{q}$, and $b\bar{s}$ beyond the ground-state 
($0^{-+}$ and $1^{--}$) doublet.  

The gross structure of the spectra of heavy-light mesons is rather 
well understood, from a combination of potential-inspired intuition 
and heavy-quark effective theory.  The fine structure of the spectra 
is not an unambiguous prediction of HQET; thus, experimental results 
may provide some surprises and some new insights.  To give an example, 
the separation of the centroids of the $j_{q}=3/2$ ($1^{++}$ and 
$2^{++}$) and $j_{q}=1/2$ ($0^{++}$ and $1^{++}$) doublets is not a 
robust prediction of the heavy-quark theory \cite{jlight}.  We believe 
that HQET and the chiral quark model do give us the tools we need to 
describe the strong-interaction transitions among mesonic states with 
precision, as described by Estia Eichten in his talk at \hq\ \cite{eje}. 
The $j_{q}=3/2$ states are expected to be narrow; the charmed 
states are thoroughly known, and the beauty states are under 
investigation in a number of experiments.  Franz Muheim \cite{lepbd} 
showed what the LEP experiments have been able to achieve in analyses 
that rely to varying degrees on the predictions \cite{ehq} of 
heavy-quark effective theory.  The experimental observations are in 
broad agreement with HQET, and indicate that a significant fraction 
of $B$ mesons (25 to 40\%) are produced through the $p$-wave $B^{**}$ 
states.  This conclusion holds great interest for the same-side 
tagging of $B$ flavor for studies of $CP$ violation.

The $j_{q}=1/2$ levels have not been established yet.  They are 
expected to be broad, but theorists are only beginning to endow that 
label with a numerical meaning.  At \hq, Jorge Rodriguez \cite{cleocb} 
presented \textsc{cleo}'s evidence for the $D_{1}^{*}(j_{q}=1/2)$ state 
near $2461\mevcc$, with a total width of about 
$200\hbox{ -- }400\mev$.  This seems considerably broader than the 
theoretical expectation \cite{eje,goity} that $\Gamma(0^{+} 
\rightarrow 0^{-}\pi) \approx \Gamma(1^{+} \rightarrow 1^{-}\pi) 
\approx 85\mev$.  In the $B$ system, an L3 analysis reported by 
Muheim \cite{lepbd} suggests that the $B_{1}^{*}$ has a mass of 
$5675 \pm 12 \pm 4\mevcc$ and a width of $78 \pm 28 \pm 15\mev$, in 
the range suggested by chiral-quark-model calculations.  We can 
expect both theoretical and experimental progress by the time of 
\textit{HQ2000.} 
 
The \textsc{delphi} Collaboration \cite{narrowdelphi} recently 
reported an excess of events in the $D^{*+}\pi^{+}\pi^{-}$ mass 
spectrum at a mass of $2637 \pm 2 \pm 6\mev$, which is consistent with 
expectations for the first radial excitation of the $D^{*}$ meson.  
The width of the excess is quite small, consistent with their 
experimental resolution.  Both \textsc{opal} (see Muheim, 
Ref.~\cite{lepbd}) and \textsc{cleo} (see Rodriguez, 
Ref.~\cite{cleocb}) have presented upper limits that are inconsistent 
with the \textsc{delphi} observation.

\subsection*{Charmed Baryons}
In SU(4)$_{\mathrm{flavor}}$ symmetry, the ground-state baryons 
include 20 $1/2^{+}$ states, of which 12 contain one or more charmed 
quarks, and 20 $3/2^{+}$ states, of which 10 contain one or more 
charmed quarks.  All nine of the singly-charmed $1/2^{+}$ states and 
four of the six singly-charmed $3/2^{+}$ states have been observed.  
Only the $\Sigma^{*+}_{c}$ and $\Omega^{*}_{c}$ remain undetected.  
Sajjad Alam \cite{cleocharm} reviewed \textsc{cleo}'s extensive 
contributions to charmed-baryon spectroscopy, while Eric Vaandering 
\cite{focusdal} summarized the achievements of Fermilab Experiment 
E687 and the promise of its successor, \textsc{focus.}  The multiply 
charmed baryons (and, more generally, heavy-heavy-light baryons) 
remain tempting experimental targets, in part for the analogy in HQET 
to heavy-light mesons \cite{eje}.

\subsection*{Charmonium Spectroscopy}
Todd Pedlar reported on Fermilab experiment E835 \cite{e835}, the 
study of charmonium spectroscopy in resonant $\bar{p}p$ annihilations.  
E835 is conducted in what we think of as the Fermi National 
\textit{De}celerator Laboratory, when the Antiproton Accumulator 
decelerates antiprotons from $8.9\gevc$ to the momenta required for 
resonant formation of $c\bar{c}$ states, in the range $4\hbox{ -- 
}7\gevc$.  E835 reports the first evidence for formation of the 
$1^{3}\mathrm{P}_{0}$ state $\chi_{0}$ in $\bar{p}p$ annihilations.  
The resonance parameters, $M(\chi_{0}) = (3415^{+2.1}_{-1.7})\mevcc$ 
and $\Gamma(\chi_{0}) = (13.9^{+5.3}_{-3.9})\mev$, are in good agreement 
with Particle Data Group averages \cite{pdg}.  The $\bar{p}p$ 
branching fraction, while consistent with the pre\"{e}xisting upper bound, 
is tantalizingly large:
\begin{equation}
	B(\chi_{0} \rightarrow \bar{p}p) = (4.24^{+0.96}_{-0.70} \pm 1.16) 
	\times 10^{-4}.
	\label{eq:chi0}
\end{equation}
The BES Collaboration has just published the first determination of 
this branching fraction in $e^{+}e^{-}$ collisions at the 
$\psi(2\mathrm{S})$ \cite{bes}; their value is $(1.59 \pm 0.43 \pm 
0.53) \times 10^{-4}$.

Despite assiduous efforts, the E835 Collaboration has not been able 
to find any sign of the expected pseudoscalar radial excitation 
$\eta_{c}^{\prime}\;(2^{1}\mathrm{S}_{0})$.  The old Crystal Ball 
claim of a state at $3594\mevcc$, which was too distant from the 
$\psi^{\prime}$ for theoretical comfort, is rather decisively ruled 
out, but it is somewhat maddening that the real $\eta_{c}^{\prime}$ 
has not shown itself.  What are we missing?
\subsection*{Production of Heavy Flavors}
A wealth of information about the production of heavy quarks was 
presented at \hq.  Fred Olness \cite{fredo} presented an excellent 
summary of the outstanding theoretical and experimental issues.  A 
key point is that heavy-quark production processes are both 
challenging and interesting for our understanding of perturbative QCD 
because they typically involve two large scales: the heavy-quark 
mass, or threshold, and the typical large-momentum scale that 
encourages the application of perturbative techniques.  Once we have 
reliable predictions for the kinematic distributions of the heavy quarks, we need to 
understand how those are reflected in the kinematic distributions of 
the hadrons that contain those quarks.  The current state of 
understanding in the Lund string-fragmentation approach was reported 
to \hq\ by Emanuel  Norrbin \cite{norrbin}.  Fermilab Experiment E791's 
new measurements of the production asymmetries $A \equiv (\sigma_{X} - 
\sigma_{\bar{X}}) / (\sigma_{X} + \sigma_{\bar{X}})$ for $D^{\pm}$, 
$D_{s}^{\pm}$, and $\Lambda_{c}$ baryons and antibaryons, presented 
by Kevin Stenson \cite{e791prod}, offer empirical insight into the 
fragmentation process.  The \textsc{hermes} experiment at 
\textsc{hera} has completed three years of successful data taking 
that includes a measurement of the cross section for $J/\psi$ 
photoproduction and a clear open charm signal \cite{hermes}, still 
under analysis.  We also look forward to the operation of the 
\textsc{compass} experiment at CERN \cite{compass}.

Collider data on charmonium production have forced us to look more 
broadly for the right physical picture of the process than the 
color-singlet mechanism in which the observed charmonium state has the 
quantum numbers of the produced $c\bar{c}$ pair.  Since the discovery 
by the CDF Collaboration \cite{cdfonium} that the direct production of both $J/\psi$ 
and $\psi^{\prime}$ occur at some fifty times the color-singlet--model 
rate, our attention has been drawn to a color-octet mechanism in which 
the produced $c\bar{c}$ pair evolves into a color-singlet hadron by 
emitting a soft gluon.  For this reason, it is sometimes called the 
color-evaporation mechanism.  At \hq, Andrzej Zieminski \cite{tevonium} 
presented recent results from the D\O\ Collaboration that test the 
color-octet picture of $J/\psi$ production into the regime of large 
rapidity and small transverse momentum.  Within uncertainties, the 
color-octet model describes the pseudorapidity dependence of $J/\psi$ 
production at all angles.  An interesting production-mechanism 
diagnostic is the polarization of the produced charmonium state.  In a 
thorough report on charm and beauty production in CERN Experiment 
WA92, Dario Barberis \cite{wa92} showed that in 350-GeV$\!/\!c$ 
$\pi^{-}$A collisions, the $J/\psi$ polarization is small, in 
agreement with the color-octet picture.  Ting-Hua Chang \cite{nusea} 
similarly reported that in 800-GeV$\!/\!c$ $p$Cu collisions, the 
\textsc{nusea} Collaboration (Fermilab Experiment E866) observes no 
polarization in the $J/\psi$ decay angular distribution integrated 
over all production angles (represented by $x_{\mathrm{F}}$).

Models for charmonium production need also to confront the 
photoproduction data shown by Beate Naroska \cite{herahf} in her 
summary of heavy-flavor work at H1 and \textsc{zeus.}  The 
\textsc{hera} measurements, taken in a kinematic regime where 
diffraction dominates, are described very well by a 
next-to-leading-order color singlet model, and do not demand a 
significant color-octet contribution.  The link between 
nonperturbative parameters set by CDF data and the consequences for 
diffractive production at \textsc{hera} involves some subtleties that 
need further work to resolve.

In heavy-ion collisions, charmonium suppression---beyond the normal 
nuclear absorption long observed in hadron--nucleus and light-ion 
collisions---has been predicted as a diagnostic for the creation of a 
quark-gluon plasma, \ie, a deconfined state of hadronic matter.  
A sudden drop in the $J/\psi$ yield as a function of the product of 
target and projectile mass numbers has been observed in Pb--Pb 
collisions by the NA50 Collaboration at CERN, as we heard in the 
heavy-ion summary by Carlos Louren\c{c}o \cite{hfhi}.  The effect is 
large: the measured point lies some $4.5\sigma$ below the 
extrapolation from smaller values of the mass product.  It is clearly 
a tantalizing hint.

The CCFR Collaboration at Fermilab has used the production of one 
heavy quark, charm, to investigate the population of another heavy 
quark, strange, in the nucleon.  Todd Adams \cite{nudis} presented the 
results of their next-to-leading-order fits to the charm production 
cross section in $(\nu_{\mu}, \bar{\nu}_{\mu})N$ deeply inelastic 
scattering.  They confirm the expectation (and the conclusion of 
earlier experiments) that the nucleon sea is not SU(3)-symmetric, and 
find that the inferred value of the charm quark mass is analysis 
dependent.  The successor experiment, NuTeV, has an improved data 
sample that will permit more incisive analyses.

\section*{Strange-Particle Decays}

\subsection*{Search for Direct {\boldmath$CP$} Violation}
$CP$ violation can arise from a small impurity in the mass 
eigenstates of the $K^{0}$--$\bar{K}^{0}$ complex, or from a 
``direct'' $CP$-violating contribution to the decay amplitudes, or 
from interference between the two.  If 
$CPT$ is a good symmetry, the mass eigenstates can be written as
 \begin{equation}
 	\ket{K_{S}} = p\ket{K^{0}} + q\ket{\bar{K}^{0}}, \qquad
 	\ket{K_{L}} = p\ket{K^{0}} - q\ket{\bar{K}^{0}}.
 	\label{eq:Keigen}
 \end{equation} 
If $CP$ invariance held, we would have $q = p = 1/\sqrt{2}$, so that 
$K_{S}$ would be $CP$-even and $K_{L}$ $CP$-odd.  In a convenient 
phase convention \cite{wwstein}, we can express $CP$ violation in 
$K^{0}$--$\bar{K}^{0}$ mixing through the parameter $\varepsilon$ 
($|\varepsilon|\approx 2.28 \times 10^{-3}$, with a phase near 
$45^{\circ}$),
\begin{equation}
	\frac{p}{q} = \frac{(1 + \varepsilon)}{(1 - \varepsilon)}.
	\label{eq:eps}
\end{equation}

$CP$ violation in the decay amplitudes gives rise to an inequality---a 
phase difference---in the amplitudes $A(K^{0}\rightarrow \pi\pi(I))$ 
and $A(\bar{K}^{0}\rightarrow \pi\pi(I))$, where $I$ is the isospin of 
the $\pi\pi$ system.  It is conventional to express the $CP$-violating 
observables in terms of the parameters $\varepsilon$ and 
$\varepsilon^{\prime}$, the latter defined through
\begin{eqnarray}
	\eta_{+-} \equiv & {\displaystyle \frac{A(K_{L} \rightarrow \pi^{+}\pi^{-})}
	{A(K_{S} \rightarrow \pi^{+}\pi^{-})}} & = \varepsilon + 
	\varepsilon^{\prime},
	\nonumber  \\
	\eta_{00} \equiv & {\displaystyle \frac{A(K_{L} \rightarrow \pi^{0}\pi^{0})}
	{A(K_{S} \rightarrow \pi^{0}\pi^{0})}} & = \varepsilon - 2 
	\varepsilon^{\prime}.
	\label{eq:epsprime}
\end{eqnarray}
The observable $|\eta_{+-}|^{2}/|\eta_{00}|^{2} \approx 1 + 
6\Real{\varepsilon^{\prime}/\varepsilon}$ is very close to unity.  In the 
electroweak theory, a tiny deviation from one arises from the phase in 
the quark mixing matrix.  In the electroweak theory, it seems most 
likely that $\varepsilon^{\prime}/\varepsilon$ should be on the order 
of $10^{-3}$, or perhaps smaller \cite{bbl}.

Published measurements of $\varepsilon^{\prime}/\varepsilon$ have 
already reached a remarkable level of precision, with E731 at Fermilab 
reporting $(0.74 \pm 0.52 \pm 0.29) \times 10^{-3}$ \cite{e731} and 
the NA31 experiment at CERN quoting $(2.3 \pm 0.65)\times 10^{-3}$ 
\cite{na31}.  If we take both of these beautiful results at face 
value, then both the existence and perforce the magnitude of a direct 
$CP$-violating amplitude remain in doubt.  Three experiments now in 
progress aim at a precision that will settle the issue.  NA48 at CERN 
and KTeV at Fermilab have already logged very significant data sets.  
Augusto Ceccucci reported \cite{na48} that NA48 anticipates a result 
based on their 1997 data in time for the winter conferences.  The 
statistical uncertainty on $\Real{\varepsilon^{\prime}/\varepsilon}$ 
should be around $(4\hbox{ -- }5)\times 10^{-4}$, and the systematic 
error should be still smaller.  According to Mike Arenton 
\cite{mikea}, KTeV is nearing a final result based on 20\% of the data 
they recorded in 1996 and 1997.  They expect a statistical uncertainty 
of about $3\times 10^{-4}$, and are currently studying systematic 
effects.  And we learned from G.  Bencivenni that when the $\phi$ 
factory DA$\Phi$NE operates at full luminosity, the \textsc{kloe} 
experiment should be able to probe 
$\Real{\varepsilon^{\prime}/\varepsilon}$ to $10^{-4}$ \cite{dafne}.

\subsection*{{\boldmath$CP$} Violation in Hyperon Decay?}
To understand the origin of $CP$ violation, it is a matter of urgent 
interest to find $CP$-violating phenomena outside the neutral-kaon 
system.  A natural place to look is in the decays of other strange 
particles, notably the hyperons of the baryon octet.  \textsc{HyperCP} 
at Fermilab, described at \hq\ by Cat James \cite{cat}, is the first 
experiment dedicated to the search for $CP$ violation in hyperon 
decay.  \textsc{HyperCP} uses a high-rate spectrometer to compare the 
decay chains
\dk{\Xi^{-}}{\Lambda \pi^{-}}{p\pi^{-}}
and
\dk{\bar{\Xi}^{+}}{\bar{\Lambda} \pi^{+}}{\bar{p}\pi^{+}.}
The decay angular distribution of the proton in the $\Lambda$ rest 
frame,
\begin{equation}
	\frac{dN}{d\cos\theta} = \cfrac{1}{2}(1+ 
	\alpha_{\Lambda}\alpha_{\Xi}\cos\theta),
	\label{eq:hyperdis}
\end{equation}
where $\theta$ is the angle between the proton momentum and the 
$\Lambda$ polarization vector,
is characterized by the asymmetry parameters of the sequential hyperon 
decays.  \textsc{HyperCP} aims to measure the $CP$-violating asymmetry,
\begin{equation}
	{\mathcal A} = \frac{\alpha_{\Lambda}\alpha_{\Xi} - 
	\alpha_{\bar{\Lambda}}\alpha_{\bar{\Xi}}} {\alpha_{\Lambda}\alpha_{\Xi} + 
	\alpha_{\bar{\Lambda}}\alpha_{\bar{\Xi}}} \approx A_{\Lambda} + 
	A_{\Xi} \;\; ,
	\label{eq:hyperasym}
\end{equation}
with a sensitivity of one part in $10^{4}$, adding data from the 1999 
run to the data in hand from 1997.  Published predictions 
for the joint asymmetry range from about $10^{-5}$ to a few times 
$10^{-4}$, whereas the superweak model predicts no asymmetry.

\subsection*{Direct Observation of {\boldmath$T$}-Violation}
Perhaps the most satisfying new result presented at \hq\ was the KTeV 
observation \cite{mikea} of a time-reversal--violating asymmetry in 
the rare decay $K_{L} \rightarrow \pi^{+}\pi^{-}e^{+}e^{-}$.  The best 
evidence that the $\pi^{+}\pi^{-}e^{+}e^{-}$ mode qualifies as rare is 
that it had not been reported until this year \cite{firstobs}.  An 
analysis of about 60\%\ of KTeV's 1997 data set now allows a precise 
determination of the branching fraction as $(3.32 \pm 0.14 \pm 0.28) 
\times 10^{-7}$.  The interest in this decay mode derives from the 
fact that the underlying process, $K_{L} \rightarrow \pi^{+}\pi^{-} 
\gamma$, proceeds through both $CP$-conserving and $CP$-violating 
mechanisms.  A Bremsstrahlung component is associated with the 
$CP$-violating $K_{L} \rightarrow \pi^{+}\pi^{-}$ decay, while a 
direct-emission component arises from a $CP$-conserving M1 transition.  
The interference between amplitudes with different $CP$ properties can 
lead to a $CP$-violating effect in the photon polarization.  The 
$K_{L} \rightarrow \pi^{+}\pi^{-}e^{+}e^{-}$ channel, which 
represents the internal conversion of the photon to an 
electron-positron pair, analyzes the 
virtual photon polarization through the orientation of the 
$e^{+}e^{-}$ plane relative to the $\pi^{+}\pi^{-}$ plane \cite{sacha}.  

To be explicit, let $\hat{\mathbf{n}}_{\pi} = (\mathbf{p}_{\pi^{+}} 
\times \mathbf{p}_{\pi^{-}}) / |\mathbf{p}_{\pi^{+}} \times 
\mathbf{p}_{\pi^{-}}|$ be the normal to the pion plane and 
$\hat{\mathbf{n}}_{e} = (\mathbf{p}_{e^{+}} \times \mathbf{p}_{e^{-}}) 
/ |\mathbf{p}_{e^{+}} \times \mathbf{p}_{e^{-}}|$ be the normal to the 
electron plane, and define the azimuthal angle $\varphi$ through 
$\cos\varphi = \hat{\mathbf{n}}_{\pi} \cdot \hat{\mathbf{n}}_{e}$.  
The decay angular distribution is of the form $d\Gamma/d\varphi = 
\Gamma_{1}\cos^{2}\varphi + \Gamma_{2}\sin^{2}\varphi + 
\Gamma_{3}\sin\varphi \cos\varphi$.  We can express $\sin\varphi = 
(\hat{\mathbf{n}}_{\pi} \times \hat{\mathbf{n}}_{e}) \cdot 
\hat{\mathbf{z}}$, where $\hat{\mathbf{z}} = (\mathbf{p}_{\pi^{+}} + 
\mathbf{p}_{\pi^{-}})/|\mathbf{p}_{\pi^{+}} + \mathbf{p}_{\pi^{-}}|$.  
Under the action of the charge conjugation operator $C$, we have 
$\mathbf{p}_{\pi^{\pm}} \rightarrow \mathbf{p}_{\pi^{\mp}}$ and 
$\mathbf{p}_{e^{\pm}} \rightarrow \mathbf{p}_{e^{\mp}}$, so that 
$\hat{\mathbf{n}}_{\pi} \rightarrow - \hat{\mathbf{n}}_{\pi}$, 
$\hat{\mathbf{n}}_{e} \rightarrow - \hat{\mathbf{n}}_{e}$, and 
$\hat{\mathbf{z}} \rightarrow \hat{\mathbf{z}}$.  Either the parity 
operator $P$ or the time-reversal operator $T$ takes 
$\mathbf{p}_{\pi^{\pm}} \rightarrow -\mathbf{p}_{\pi^{\pm}}$ and 
$\mathbf{p}_{e^{\pm}} \rightarrow -\mathbf{p}_{e^{\pm}}$, so that 
$\hat{\mathbf{n}}_{\pi} \rightarrow \hat{\mathbf{n}}_{\pi}$, 
$\hat{\mathbf{n}}_{e} \rightarrow \hat{\mathbf{n}}_{e}$, and 
$\hat{\mathbf{z}} \rightarrow - \hat{\mathbf{z}}$.  Accordingly, $P$ 
and $T$ take $\sin\varphi \rightarrow - \sin\varphi$ and $\cos\varphi 
\rightarrow \cos\varphi$, while $C$ leaves both $\sin\varphi$ and 
$\cos\varphi$ unchanged.  The presence of a $\sin\varphi \cos\varphi$ 
term in the decay angular distribution, \ie, of a nonzero value of 
$\Gamma_{3}$, is direct evidence for time-reversal noninvariance and, 
since $C$ leaves the decay angular distribution unchanged, for $CP$ 
violation.  Indirect $CP$ violation---the same physics that produces a 
nonzero value of $\varepsilon$---induces a $T$-violating asymmetry in 
the decay-angular distribution whose size is determined by 
$\Imag{\varepsilon}$.  The effect is large, because the 
$CP$-violating contribution to the $K_{L} \rightarrow 
\pi^{+}\pi^{-}e^{+}e^{-}$ decay amplitude occurs at a lower order in 
chiral perturbation theory than the $CP$-conserving contribution. Sehgal and 
Wanninger \cite{sehgal} have computed a forward-backward asymmetry $A 
= (14.3 \pm 1.3) \%$.

KTeV's full 1997 data set leads to a sample of $1811 
\pm 42$ events, enough to study the decay angular distributions in 
detail.  The preliminary asymmetry presented at \hq\ is
\begin{equation}
	A = (13.5 \pm 2.5_{\mathrm{stat}} \pm 3.0_{\mathrm{sys}})\% ,
	\label{eq:KTeVasym}
\end{equation}
which represents direct evidence for a violation of time-reversal 
symmetry.  This is the largest particle-antiparticle asymmetry so far 
observed.  The measured value is in good agreement with theoretical 
expectations.

During the week of \hq, the CPLEAR Collaboration at CERN reported on 
the first observation of time-reversal symmetry violation through a 
comparison of the probabilities for the transformations $\bar{K}^{0} 
\leftrightarrow K^{0}$ as a function of the neutral-kaon proper time 
\cite{CPLEAR}.  
In their experiment, the strangeness of the neutral kaon at the moment 
of its creation, $t=0$, was tagged by observing the kaon charge in the 
formation reaction $\bar{p}p \rightarrow K^{\pm}\pi^{\mp}(K^{0}, 
\bar{K}^{0})$ at rest, while the strangeness of the neutral kaon at 
the time of its semileptonic decay, $t=\tau$, was tagged by the charge 
of the final-state lepton.  The time-average decay-rate asymmetry, 
measured over the interval $1\times\tau_{s} < \tau < 20\times \tau_{s}$, is
\begin{equation}
	\left\langle \frac{\Gamma({\bar{K}^{0}}|_{0} \rightarrow 
	{e^{+}\pi^{-}\nu}|_{\tau}) - \Gamma({K^{0}}|_{0} \rightarrow 
	{e^{-}\pi^{+}\bar{\nu}}|_{\tau})}{\Gamma({\bar{K}^{0}}|_{0} \rightarrow 
	{e^{+}\pi^{-}\nu}|_{\tau}) + \Gamma({K^{0}}|_{0} \rightarrow 
	{e^{-}\pi^{+}\bar{\nu}}|_{\tau})}\right\rangle =
	(6.6 \pm 1.3_{\mathrm{stat}}\pm 1.0_{\mathrm{sys}}) \times 10^{-3}.
	\label{eq:CPLEAR}
\end{equation}
This asymmetry is a direct manifestation of $T$-violation.  If $CPT$ 
is a good symmetry in semileptonic decays and the $\Delta S=\Delta Q$ 
rule is exact, then the observed asymmetry \eqn{eq:CPLEAR} is 
identical to
\begin{equation}
	\frac{{\mathcal P}(\bar{K}^{0} \rightarrow K^{0}) - {\mathcal P}(K^{0} \rightarrow \bar{K}^{0})}
	{{\mathcal P}(\bar{K}^{0} \rightarrow K^{0}) + {\mathcal P}(K^{0} \rightarrow 
	\bar{K}^{0})} ,
	\label{eq:probasym}
\end{equation}
where ${\mathcal P}$ is a probability for strangeness oscillation.  
The observed result is in good agreement with the theoretical 
expectation, $4\Real{\varepsilon} = (6.63 \pm 0.06) \times 10^{-3}$.

These two new results confirm our expectation that time-reversal 
invariance is violated in neutral-kaon decays, as must be the case 
if $CPT$ holds and $CP$ is not respected.  In quantitative terms, the 
newly observed $T$-violations occur at just the level required to 
compensate for the $CP$ violation known since 1964 to occur in the 
decay $K_{L} \rightarrow \pi^{+}\pi^{-}$.

\subsection*{Plenty of Nothing}
One of the most beautiful results of the year past was the observation 
by Brookhaven Experiment 787 \cite{e787} of a single, very clean, example of the 
decay
\begin{equation}
	K^{+} \rightarrow \pi^{+}\nu \bar{\nu},
	\label{eq:kpinunu}
\end{equation}
corresponding to a branching fraction of $(4.2^{+9.7}_{-3.5})\times 
10^{-10}$ that is consistent with the standard-model expectation, $0.6 
\times 10^{-10} \le B(K^{+} \rightarrow \pi^{+}\nu \bar{\nu}) \le 1.5 
\times 10^{-10}$.  What is most impressive to me is not the one 
beautiful candidate event, but the extremely low level of background: 
the event occurs on an empty field.  In the report presented to \hq\ 
by Steve Kettell \cite{e787hq}, we learned that preliminary indications from the 
analysis of 1995--1997 data are that the background rejection is three 
times better, with an increased acceptance.  Over the next two years, 
the E787 sensitivity should provide a thorough survey of the 
standard-model regime.  Brookhaven proposal 949 \cite{e949} would increase the 
sensitivity to $10^{-11}$, and the CKM proposal at Fermilab 
\cite{ckmprop} aims at a sensitivity of $10^{-12}$.  As long as the 
experimental sensitivity fell far short of standard-model 
expectations, the principal interest of searching for 
$K^{+} \rightarrow \pi^{+}\nu \bar{\nu}$ was to probe non--standard-model 
physics.  With detection achieved near the band of standard-model 
predictions, the branching ratio takes on additional importance as 
a determination of $|V_{td}|$.

A little less nothing has been achieved by KTeV in its search for the 
companion process, $K_{L} \rightarrow \pi^{0}\nu \bar{\nu}$.  With a 
small expected background of 0.12 event in the signal region for the 
Dalitz-decay final state $(e^{+}e^{-}\gamma)\nu \bar{\nu}$, they 
observe no events, and can quote an upper limit $B(K_{L} \rightarrow 
\pi^{0}\nu \bar{\nu}) < 4.9 \times 10^{-7}$ at 90\% confidence level 
\cite{ez}.  Dedicated experiments to measure the $K_{L} \rightarrow 
\pi^{0}\nu \bar{\nu}$ branching fraction are being planned at Fermilab 
and Brookhaven \cite{kami,e926}.  These would provide an unambiguous 
determination of $\Imag{V_{td}}$.

\subsection*{Rarest of Them All}
Two formerly rare decays are now being studied with impressive 
statistical power.  KTeV has detected 275 candidates for the decay 
$\pi^{0} \rightarrow e^{+}e^{-}$ over an expected background of 21 
events, for a preliminary branching fraction $B(\pi^{0} \rightarrow 
e^{+}e^{-}) = (6.09 \pm 0.40_{\mathrm{stat}} \pm 0.23_{\mathrm{sys}}) 
\times 10^{-8}$ \cite{ez}.  This is in good agreement with theoretical 
expectations \cite{qandj}.  Brookhaven Experiment E871 now has 
accumulated over 6200 candidates for the decay $K_{L} \rightarrow 
\mu^{+} \mu^{-}$ \cite{e871}.  That, too, is in good agreement with 
theoretical expectations \cite{german}.  They also hold the record 
sensitivity for the forbidden decay $K_{L} \rightarrow \mu^{\pm} 
e^{\mp}$, and can quote an upper limit on the branching fraction, 
$B(K_{L} \rightarrow \mu^{\pm} e^{\mp}) < 4.8 \times 10^{-12}$.

E871 holds the further distinction of measuring the smallest branching 
fraction of them all, with their observation of four candidates for 
the decay $K_{L} \rightarrow e^{+}e^{-}$ \cite{e871}.  These four 
events lead to a branching fraction $B(K_{L} \rightarrow e^{+}e^{-}) = 
(8.7^{+5.7}_{-4.1})\times 10^{-12}$, in close agreement with modern 
calculations based on chiral perturbation theory \cite{german}.  This 
is a very impressive achievement indeed.

Since 1995, Brookhaven experiment E865 has collected data on many rare 
and formerly rare decays \cite{e865}.  Among their targets is the 
lepton-flavor-violating decay $K^{+} \rightarrow 
\pi^{+}\mu^{+}e^{-}$, for which they have already set a 90\% CL upper 
limit of $2.1 \times 10^{-10}$.  The projected single-event 
sensitivity is $10^{-11}$.

\subsection*{Spinoffs}
Although the focus of KTeV is the precision study of neutral kaon 
decays, the KTeV spectrometer is also well-matched to a number of 
other important physics goals.  Doug Jensen \cite{ktevY} presented a preliminary 
measurement of the branching fraction $B(\Xi^{0}\rightarrow 
\Sigma^{+}e^{-}\bar{\nu}_{e})$ based on $153 \pm 13$ events that fit 
the pattern
\dk{\Xi^{0}}{\Sigma^{+}e^{-}\bar{\nu}_{e}}{p\pi^{0},}
upon a background of $6 \pm 2$ events.  The preliminary branching 
fraction is $(2.5 \pm 0.2 \pm 0.3)\times 10^{-4}$, in excellent accord 
with the theoretical expectation of $2.61 \times 10^{-4}$.

In a similar spirit, the \textsc{selex} experiment, which is mainly 
concerned with the study of charmed particles produced in a 
$\Sigma^{-}$ beam, has obtained interesting new results on hyperon 
properties \cite{selexY}.  They have determined the $\Sigma^{-}$ 
charge radius to be $\langle r^{2}\rangle = (0.60 \pm 0.08_{\mathrm{stat}} \pm 
0.08_{\mathrm{sys}})\fermi^{2}$, measured the $\Sigma^{-}p$ total cross section 
to be about $36\mb$ at $600\gevc$, and set a new upper bound on the $U$-spin--forbidden 
transition rate, $\Gamma(\Sigma(1385)^{-} \rightarrow 
\Sigma^{-}\gamma) < 12\kev$ at 95\% CL.

\section*{Weak Interactions \\ of Charm and Beauty}
\subsection*{Lifetimes}
The lifetimes of hadrons containing $c$ and $b$ quarks have important 
engineering value and give us insight into the interplay between the 
strong and weak interactions.  With the development of the heavy-quark 
expansion, theorists now have well-defined expectations for the 
hierarchy of $b$-hadron lifetimes that high-precision data can 
confront.  At \hq, Harry Cheung \cite{harryc} presented a survey of 
recent progress in lifetime measurements.  Speaking for the E791 
Collaboration at Fermilab, Nader Copty \cite{e791} presented a new 
precise measurement of the $D_{s}$ lifetime, $\tau_{D_{s}} = 0.518 \pm 
0.014 \pm 0.007\ps$.  This is considerably larger than the Particle 
Data Group average, $\langle \tau_{D_{s}} \rangle_{\mathrm{PDG}} = 
0.467 \pm 0.017\ps$ \cite{pdg}.  Combined with the PDG average 
lifetime for the $D^{0}$, $\langle \tau_{D^{0}}\rangle_{\mathrm{PDG}} 
= 0.415 \pm 0.004\ps$, the E791 value gives a ratio 
$\tau_{D_{s}}/\tau_{D^{0}} = 1.25 \pm 0.04$, a six-standard-deviation 
difference from unity.  [Harry Cheung's world average, including the 
new data from E791, is $\tau_{D_{s}}/\tau_{D^{0}} = 1.193 \pm 0.027$.] 
This represents a substantial change from the PDG98 value of $1.125 
\pm 0.042$.  We expect further improvements in our knowledge of charm 
lifetimes from \textsc{cleo} analyses using the new silicon vertex 
detector and the forthcoming \textsc{focus} data described by Jonathan 
Link \cite{focushad} and Eric Vaandering \cite{focusdal}, which will 
be statistically dominant over the next few years.

The CDF Collaboration has recently used semileptonic decays to 
determine the lifetime of the $B_{c}$ meson as $\tau_{B_{c}} = 
0.46^{+0.18}_{-0.16} \pm 0.03\ps$ \cite{bsubc}, very close to the 
value expected in the spectator picture \cite{bandb}.
\subsection*{Semileptonic decays}
The semileptonic decays of charm and beauty are of interest for the 
light they can shed on quark mixing matrix elements and on the 
dynamics embodied in hadronic form factors.  We want to both test and 
exploit the predictions of heavy-quark effective theory for the 
behavior of form factors and for the connection between $D$ and $B$ 
decays.  

The study of semileptonic $B$ decays constrains the parameters 
$|V_{cb}|$ and $|V_{ub}|$ that will be crucial for interpreting 
$CP$-violating effects in $B$ decays.  Karl Ecklund \cite{cleosl} 
reported recent \textsc{cleo} results on the exclusive 
reconstruction of $B \rightarrow D\ell\nu$ and $B \rightarrow 
\rho\ell\nu$, and presented a new moment analysis of inclusive 
semileptonic $B$ decays that may reduce the uncertainties in 
extracting $V_{cb}$.

Fermilab Experiment E791 has made important strides in the study of 
form factor ratios in the decays $D^{+} \rightarrow 
\bar{K}^{*}\ell^{+}\nu_{\ell}$ and $D_{s}^{+} \rightarrow 
\phi\ell^{+}\nu_{\ell}$.  Daniel Mihalcea \cite{e791sl} showed the 
evolution of these measurements, which now offer a worthy challenge 
for theory based on lattice QCD.  Fermilab Experiment E687 has made 
competitive measurements of semileptonic form factors in the past.  
Will Johns \cite{focussl} reviewed these contributions and 
demonstrated that the successor experiment, \textsc{focus}, will yield 
thirty to forty times the number of events used in the E687 
semileptonic analyses.  The gigantic event sample raises the prospect 
of high-precision studies of Cabibbo-suppressed semileptonic decays of charm.

\subsection*{More Promises and Prospects}
The \textsc{selex} experiment at Fermilab is a new spectrometer that 
took data in 1996--1997 with 600-GeV$\!/\!c$ $\Sigma^{-}$ and 
$\pi^{-}$ and 540-GeV$\!/\!c$ proton beams.  They are just beginning 
to produce preliminary results on their large sample of charm decays.  
Alex Kushnirenko \cite{selexc} reported the first observation of the 
Cabibbo-suppessed decay $\Xi_{c}^{+} \rightarrow pK^{-}\pi^{+}$.

Mitsuhiro Nakamura \cite{nucs} presented new limits on $\nu_{\mu} 
\rightarrow \nu_{\tau}$ oscillations from the emulsion experiment 
\textsc{chorus} at CERN.  They have been able to move the exclusion 
plot (in the $\Delta m^{2}\hbox{ -- }\sin{2\theta}$ plane) near 
$\sin{2\theta} = 10^{-3}$, a significant increase in sensitivity over 
previous experiments.  Of particular interest to the heavy-quark 
community are the remarkable advances that have been achieved in 
automated emulsion scanning.  Another order of magnitude in scanning 
power should be in hand by \textit{HQ2000.}

\subsection*{Rare decays}
Gustavo Burdman \cite{gustavo} presented an elegant summary of the 
potential of rare $K$, $D$, and $B$ decays to probe the structure of 
the electroweak theory at one-loop level.  By looking for effects 
that derive from higher-order processes in the standard model, we may 
hope to probe momentum scales at or above the scale of electroweak 
symmetry breaking.  The theoretical art lies in identifying processes 
in which the dominant contributions are from short-distance (high 
momentum scale) processes.  

The experimental status of searches for rare decays and $CP$ violation 
in the charm system was summarized by Simon Kwan \cite{simon}.  No 
$CP$ violation has been observed, with a sensitivity of a few percent.  
The current limits on rare and forbidden $D$ decays are at the level 
of one part in $10^{5}$.  The current experimental limits on flavor 
changing neutral current processes are still orders of magnitude above 
the standard-model expectation, so there is a large window for the 
discovery of new physics.  In the immediate future, the greatest 
sensitivity to $CP$ violation in the charm sector will come from the 
$B$ factory experiments, \textsc{BaBar}, \textsc{belle}, and 
\textsc{cleo~iii.}

\section*{Summary Remarks}
It is a glorious time for heavy-quark physics.  The results presented 
at \hq\ reflect dramatic progress over the past decade and offer 
immense promise for the years ahead.  For each of the heavy 
quarks---strange, charm, and beauty---that have occupied our attention 
at this workshop, experiments in progress and under construction will 
decisively improve the quality and amount of information available to 
us.  And let us not forget the torrent of new information about the 
top quark that the next run of the Tevatron Collider will bring 
\cite{thinkshop}.  Theoretical advances make it ever clearer that we 
will be able to interpret the new experimental findings to get at the 
essence of the interactions of heavy quarks.  I look forward, with 
eager anticipation, to \textit{HQ2000} and beyond.

\section*{Acknowledgements}
It is a pleasure to thank Joel Butler and the local organizing 
committee for the stimulating and pleasant atmosphere of 
\hq. I am grateful to my scientific secretaries, Erik 
Gottschalk, Rob Kutschke, and Erik Ramberg, for providing me with 
insightful advice and timely copies of transparencies.  I thank the 
workshop staff for performing many small miracles, and Harry Cheung 
for his attention to the \hq\ \textit{Proceedings.}  Andreas 
Kronfeld, Zoltan Ligeti, and Bruce Winstein made helpful comments on 
the manuscript.


\begin{references}
\bibitem{jlight} We use $j_{q}$ ($q$ for light quark) to denote the 
vector sum of the light-quark spin and the orbital angular momentum 
between the quark and antiquark.
\end{references}
\end{document}